\documentclass[conference]{IEEEtran}
\IEEEoverridecommandlockouts
\usepackage{cite}
\usepackage{amsmath,amssymb,amsfonts}
\usepackage{algorithmic}
\usepackage{graphicx}
\usepackage{textcomp}
\usepackage{xcolor}
\usepackage{comment}
\def\BibTeX{{\rm B\kern-.05em{\sc i\kern-.025em b}\kern-.08em
    T\kern-.1667em\lower.7ex\hbox{E}\kern-.125emX}}

\makeatletter
\newcommand{\linebreakand}{%
  \end{@IEEEauthorhalign}
  \hfill\mbox{}\par
  \mbox{}\hfill\begin{@IEEEauthorhalign}
}
\makeatother

\usepackage{dblfloatfix}

\usepackage[hidelinks]{hyperref}
    
\begin{document}

\title{Towards Living Software Architecture Diagrams
}

\author{\IEEEauthorblockN{Filipe F. Correia}
\IEEEauthorblockA{\textit{Faculty of Engineering} \\
\textit{University of Porto}\\
\textit{INESCTEC}\\
Porto, Portugal \\
filipe.correia@fe.up.pt}
\and
\IEEEauthorblockN{Ricardo Ferreira}
\IEEEauthorblockA{\textit{Faculty of Engineering} \\
\textit{University of Porto}\\
\textit{INESCTEC}\\
Porto, Portugal \\
ee03195@fe.up.pt}
\and
\IEEEauthorblockN{Paulo G. G Queiroz}
\IEEEauthorblockA{\textit{Federal Rural University} \\
\textit{of the Semi-arid Region}\\
\textit{INESCTEC}\\
Porto, Portugal \\
pgabriel@ufersa.edu.br}
\linebreakand
\IEEEauthorblockN{Henrique Nunes}
\IEEEauthorblockA{\textit{Faculty of Engineering} \\
\textit{University of Porto}\\
\textit{INESCTEC}\\
Porto, Portugal \\
up201906852@edu.fe.up.pt}
\and
\IEEEauthorblockN{Matilde Barra}
\IEEEauthorblockA{\textit{Faculty of Engineering} \\
\textit{University of Porto}\\
\textit{INESCTEC}\\
Porto, Portugal \\
up201904795@edu.fe.up.pt}\\
\and
\IEEEauthorblockN{Duarte Figueiredo}
\IEEEauthorblockA{\textit{Faculty of Engineering} \\
\textit{University of Porto}\\
\textit{INESCTEC}\\
Porto, Portugal \\
up201202379@edu.fe.up.pt}
}

\maketitle

\begin{abstract}
Software architecture often consists of interconnected components dispersed across source code and other development artifacts, making visualization difficult without costly additional documentation. Although some tools can automatically generate architectural diagrams, these hardly fully reflect the architecture of the system. We propose the value of automatic architecture recovery from multiple software artifacts, combined with the ability to manually adjust recovered models and automate the recovery process. We present a general approach to achieve this and describe a tool that generates architectural diagrams for a software system by analyzing its software artifacts and unifying them into a comprehensive system representation. This representation can be manually modified while ensuring that changes are reintegrated into the diagram when it is regenerated. We argue that adopting a similar approach in other types of documentation tools is possible and can render similar benefits.
\end{abstract}


\section{Introduction}

The architecture of a software system encompasses the design decisions that shape its structure, including the major elements, their relationships, and how they align with business objectives~\cite{book:SofArcPra}. Given the profound impact of architecture on software quality, ensuring that the actual implementation aligns with design decisions is a continuous activity throughout the software development lifecycle and maintenance phases to guarantee software trustworthiness.

Diagrammatic descriptions of the software architecture can make it easier to be understood and foster collaboration within teams. Nevertheless, as with other manually generated artifacts, they tend to become outdated as the implementation evolves, especially due to a lack of documentation practices, knowledge gaps, or neglect~\cite{article:SofDocIss}. This issue can and makes software reuse, deployment, and maintenance significantly more difficult~\cite{article:ImpactOfSoftware}.

Tools to help document software have existed for a long time~\cite{knuth1983web,friendly1996design,aguiar2003minimalist}, many of them leaning on known best practices~\cite{correia2009patterns}, but their use still requires a non-trivial effort from developers to ensure the documentation is up-to-date and consistent.

The notion of \textit{Living Documentation} emerged as a reaction to the challenges of keeping the value of software documentation throughout time~\cite{book:LivDocBy}. It refers to any form of dynamic documentation that evolves continuously to reflect the current state of a system. While tools exist for automatically generating architecture diagrams~\cite{article:AFraFor}, they often have limited applicability, are complex, or lack versatility, expressiveness, or adherence to standardized notations (\textit{e.g.}, the Unified Modeling Language---UML). Furthermore, the output of these tools is usually not fully representative of what the architecture is, as the software artifacts used to generate it (\textit{e.g.}, source code) do not contain all the information needed to completely depict the architecture. As a consequence, it is common for developers to manually change a generated diagram, fine-tuning it and making it more complete and precise, by changing or adding new elements. Such a workflow can bring its own challenges, especially when the implementation changes, as generating the diagram a second time will usually imply that manual modifications are lost.

In this position paper, we argue that it is possible for tools supporting \textit{living software architecture diagrams} to achieve the best of both worlds---they can provide a complete depiction of the actual architecture while still supporting manual edits that are retained as the implementation changes and the architectural diagram is continually updated. 

In this paper, we present a general data-driven approach to designing a tool that supports these capabilities (\textit{cf.} Section~\ref{sec:approach}); then, we outline a set of specific features that such a tool can exhibit to support the general approach, and we present an early prototype of these features (\textit{cf.} Section~\ref{sec:architecture}); 
finally, we present our conclusions (\textit{cf.} Section~\ref{sec:conclusions}).

\section{Approach}
\label{sec:approach}

The approach we propose combines a data-driven process of generating a model of the software architecture with support for manual changes to render the documentation more consistent and complete. The general workflow of our approach consists of the steps described below, which can happen every time the implementation changes and a user requests a diagram update.

\begin{itemize}
    \item \textbf{Gather architectural information}: Collect architecturally-significant information from a system (\textit{e.g.}, source code, configuration files, build scripts) or from its operational context (\textit{e.g.}, log files, network communications, resource usage). Depending on the data sources used, it is possible that such information will provide a partial or imprecise view of the software architecture. The result of this step is a model of the architecture that combines information from the considered data sources.
    
    \item \textbf{Automatically reapply manual changes}: Update the model resulting from the previous step by reapplying user edits that had been made in previous versions of the diagram. Such edits will have been stored of a sequence of \textit{commands} that can be executed again at any time. They are reapplied in the same order as they have been previously made.
    
    \item \textbf{Generate architectural diagram}: The architectural model is used to generate one or more rich visual representations of the architecture, which can be embedded or linked to in project documentation. The layout of the diagram should be automatic to account for a fully data-drive process (\textit{i.e.}, when no manual changes are done).
    
    \item \textbf{Edit architectural diagram}: Users can edit the resulting diagram. This can involve making the names of diagram elements more expressive, adding or removing elements, or establish or removing connections. If manual changes are introduced, the sequence of changes is stored in such a way that it can be automatically reapplied at any a later time (see the second step, above).
\end{itemize}

\section{Infragenie}
\label{sec:architecture}

We started implementing this approach as \textit{Infragenie}, a prototype tool realized taking the three main design decisions of (a) using Docker Compose files as data source, (b) using UML component diagrams as output notation, and (c) ability to edit the resulting diagram through the PlantUML syntax\footnote{https://plantuml.com/}. 
With this tool we intend to build a proof of content that shows the viability of the approach. 

\begin{figure}[!h]
  \begin{center}
    \includegraphics[width=1\linewidth]{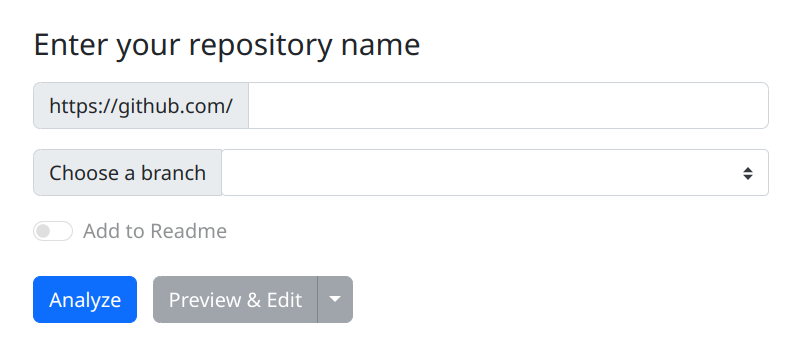}
    \caption{User interface of \textit{infragenie} main form.}
    \label{fig:infragenie-homepage} 
  \end{center}
\end{figure}

The tool was developed as a web application and is currently available at the \textit{\href{https://infragenie.eu}{infragenie.eu}} domain. The user interface of the entrypoint to the application is presented in Figure~\ref{fig:infragenie-homepage}. The user provides the address of the GitHub repository and branch to be analyzed.

The \textit{Analyze} button triggers the creation of a pull request in the specified repository, which adds a diagram image and changes the README.md file if that option is selected (\textit{cf.} Figure~\ref{fig:infragenie-homepage}). The pull request includes a direct URL to start the editor. 

\begin{figure*}[!t]
  \begin{center}
    \includegraphics[width=0.95\textwidth]{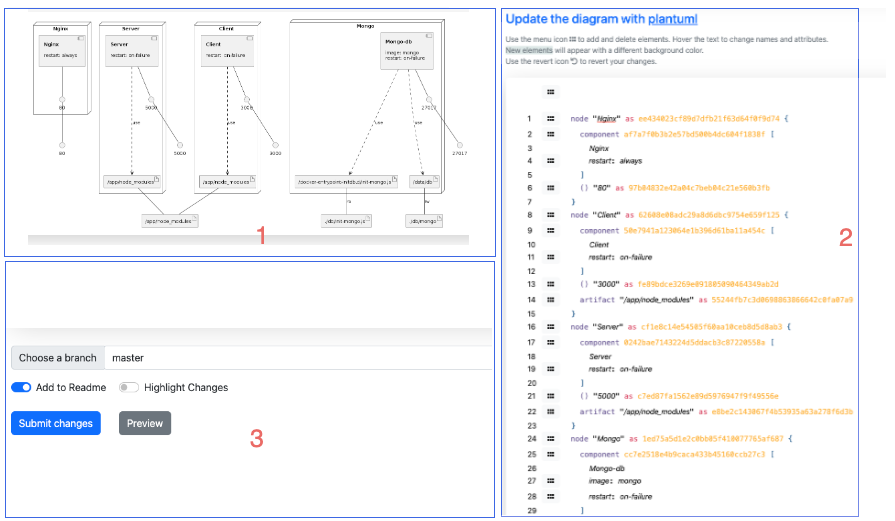}
    \caption{Infragenie: Edit page.}
    \label{fig:edit-page} 
  \end{center}
\end{figure*}

\begin{figure*}[!b]
  \begin{center}
    \includegraphics[width=0.55\textwidth]{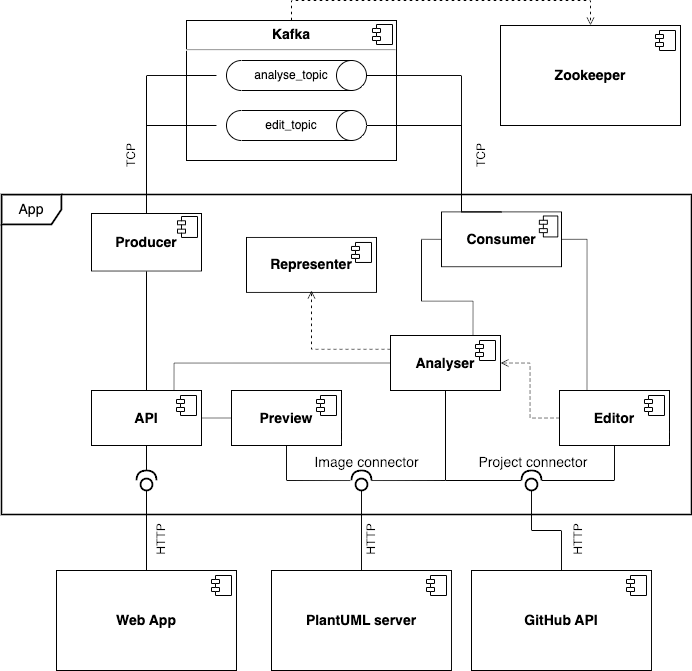}
    \caption{Prototype's conceptual high-level architecture.}
    \label{fig:conceptual-architecture} 
  \end{center}
\end{figure*}

The \textit{Preview \& Edit} button allows the user to check the diagram and edit it before creating the pull request. Editing the diagram is done by interacting with PlantUML source that generates the diagram (\textit{cf.} Figure~\ref{fig:edit-page}). To facilitate editing and help the user, we include the following features: 
\begin{itemize}
    \item To support the process of re-applying manual edits mentioned above and limit users to expressing valid diagram elements, the editor does not allow PlantUML text to be freely inserted. Instead, each line provides a context menu where the allowed changes for the PlantUML element in that line can be performed. 
    \item Every change in the diagram reveals a revert icon to facilitate the rollback of the changes;
    \item All new additions to the diagram appear with a different background;
    \item The form on the third zone of the edit page (\textit{cf.}~Figure~\ref{fig:edit-page}) has a \textit{Preview} button that updates the diagram in the first zone;
    \item We also include a feature such that these changes can be highlighted by selecting the \textit{Highlight Changes} radio button presented in part 3 of Figure~\ref{fig:edit-page}.
\end{itemize}

In combination with the preview feature, the editor allows corrections to the diagram, as well as experimenting with changes to the infrastructure and plan and understanding the impact of future changes on the architecture. When the user completes its changes, the form in the third zone of Figure~\ref{fig:edit-page} can be used to submit the changes, generating a new pull request for the project with edit event details. 

An overview of Infragenie architecture is shown in Figure~\ref{fig:conceptual-architecture}. The application is structured into \textit{Web App} (frontend) and \textit{App} (backend) modules. It relies on \textit{Kafka} to asynchronously manage requests , uses the \textit{PlantUML server} to render diagrams, and the \textit{GitHub} API to gather project information. This architecture does not require any data storage by the \textit{Backend} component, and all results are saved in the \textit{GitHub} project folder, ensuring that all resources needed for generating and recovering diagrams are available from the repository.

\section{Conclusions}
\label{sec:conclusions}

In this position paper, we introduced and approach to the development of tools for documenting a software architecture that can automatically generate and maintain up-to-date software architecture diagrams. We also present Infragenie, a prototype tool that implements our approach. This tool addresses the challenge of maintaining documentation aligned with the evolving software codebase by leveraging a combination of automated analysis and user-intervened modifications. Our approach ensures that architectural diagrams remain relevant and useful, thereby enhancing collaboration and comprehension within the development teams.

By automating the generation of architecture diagrams and allowing for manual adjustments that persist across updates, we can bridge the gap between static documentation and the dynamic nature of software development. We expect this capability will not only reduces the effort required to maintain accurate documentation but also promotes better documentation practices and architectural understanding.

\bibliographystyle{plain}
\bibliography{ref}
\end{document}